\documentclass[preprint,preprintnumbers,amsmath,amssymb]{revtex4}
\usepackage{epsfig}
\newcommand{\beq}{\begin{equation}}
\newcommand{\eeq}{\end{equation}} 
\newcommand{\beqa}{\begin{eqnarray}}
\newcommand{\eeqa}{\end{eqnarray}}
\newcommand{\ba}{\begin{array}}
\newcommand{\ea}{\end{array}}

\begin{document}

\widetext 
\title{Three-component Fermi gas with SU(3) symmetry: \\
BCS-BEC crossover in three and two dimensions} 
\author{L. Salasnich} 
\affiliation{Dipartimento di Fisica ``Galileo Galilei'' and CNISM, 
Universit\`a di Padova, \\
Via Marzolo 8, 35122 Padova, Italy\\
E-mail: luca.salasnich@unipd.it} 

\begin{abstract}
We analyze the crossover from the Bardeen-Cooper-Schrieffer (BCS) 
state of weakly bound Fermi pairs to the Bose-Einstein condensate (BEC) 
of molecular dimers for a Fermi gas made 
of neutral atoms in three hyperfine states with a 
SU(3) invariant attractive interaction. 
By solving the extended BCS equations for the 
total number of particles and the pairing gap, we calculate 
at zero temperature the pairing gap, 
the population imbalance, the condensate fraction 
and the first sound velocity of the uniform system as a function of 
the interaction strength in both three and two dimensions. 
Contrary to the three-dimensional case, 
in two dimensions the condensate fraction 
approaches the value $1$ only for an extremely large interaction strength 
and, moreover, the sound velocity gives a clear signature of the 
disappearance of one of the three hyperfine components. 
\end{abstract}

\maketitle

\section{Introduction} 

In the last years degenerate ultracold gases made of bosonic 
or fermionic atoms have been the subject of 
intense experimental and theoretical 
research \cite{book1,book2}. Among the several hot topics 
recently investigated, let us remind the expansion of a Fermi superfluid 
in the crossover from the Bardeen-Cooper-Schrieffer (BCS) 
state of weakly bound Fermi pairs to the Bose-Einstein condensate (BEC) 
of molecular dimers \cite{sigh07a,sigh07b}; the surface effects 
in the unitary Fermi gas \cite{sigh10a}; the localization 
of matter waves in optical lattices \cite{sigh10b}; 
the transition to quantum turbolence in finite-size 
superfluids \cite{sigh10c,sigh11}. 

Very recently degenerate three-component gases have been 
experimentally realized using the three lowest hyperfine states 
of $^6$Li \cite{selim,hara}. At high magnetic fields 
the scattering lengths of this three-component system are 
very close each other and the system is approximately 
SU(3) invariant. Moreover, it has been theoretically 
predicted that good $SU(N)$ invariance (with $N\leq 10$) 
can be reached with ultracold alkaline-earth 
atoms (e.g. with $^{87}$Sr atoms) \cite{wu1,wu2,ana}. 
In the past various authors \cite{various1,various2,various3,various4} 
have considered the BCS regime 
of a fermionic gas with SU(3) symmetry. In the last years 
He, Jin and Zhang \cite{cinesi} and 
Ozawa and Baym \cite{baym} have investigated the full BCS-BEC crossover 
\cite{eagles,leggett,randeria,marini,sigh08} 
of this system at zero and finite temperature in three-dimensional space. 
Recently we have calculated 
the condensate fraction and the population imbalance for this 
three-component quantum gas both in the three-dimensional case and in the  
two-dimensional one \cite{iome}. 
In this paper we review the extended BEC theory 
\cite{leggett,randeria,marini} for an atomic gas with 
three-component fermions at zero temperature \cite{cinesi,baym,iome} but 
without invoking functional integration. 
We obtain the chemical potential, 
the energy gap, the number densities and the 
condensate fraction as a function of the adimensional interaction strength. 
Finally, we calculate also the first sound velocity of the system 
both in three and two dimensions. In two dimensions we find that 
the sound velocity shows a kink at the critical strength 
(scaled binding energy) where one of the three hyperfine 
components goes to zero. 

The Lagrangian density of a dilute and ultracold 
three-component uniform Fermi gas of neutral atoms 
is given by 
\beq
{\hat {\cal L}} = 
\sum_{\alpha=R,G,B} {\hat \psi}^+_{\alpha} 
\left(i\hbar {\partial \over \partial t} 
+{\hbar^2\over 2 m}\nabla^2 + \mu \right) {\hat \psi}_{\alpha}
- g \left( {\hat \psi}^+_{R}{\hat \psi}^+_{G}
{\hat \psi}_{G}{\hat \psi}_{R} + {\hat \psi}^+_{R}{\hat \psi}^+_{B}
{\hat \psi}_{B}{\hat \psi}_{R} 
+ {\hat \psi}^+_{G}{\hat \psi}^+_{B}
{\hat \psi}_{B}{\hat \psi}_{G} \right) \; , 
\label{ham} 
\eeq
where ${\hat \psi}_{\alpha}({\bf r},t)$ is the field operator 
that destroys a fermion of component $\alpha$ 
in the position ${\bf r}$ at time $t$, 
while ${\hat \psi}_{\alpha}^+({\bf r})$ 
creates a fermion of component $\alpha$ in ${\bf r}$ at time $t$. 
To mimic QCD the three components are thought as three colors: red (R), 
green (G) and blue (B). 
The attractive inter-atomic interaction is described by a contact 
pseudo-potential of strength $g$ ($g<0$). 
The average total number of fermions is given by 
\beq 
N=\sum_{\alpha=R,G,B}
\int \langle
{\hat \psi}^+_{\alpha}({\bf r},t){\hat \psi}_{\alpha}({\bf r},t)\rangle \, 
d^3{\bf r} \; ,  
\label{numero}
\eeq
where $\langle \cdot\cdot\cdot \rangle$ is the ground-state average. 
Note that $N$ is fixed by the chemical potential $\mu$ 
which appears in Eq. (\ref{ham}). As stressed in Refs. \cite{cinesi,baym}, 
by fixing only the total chemical potential $\mu$ 
(or equivalently only the total number of atoms $N$) 
the Lagrangian (\ref{ham}) is invariant under global SU(3) rotations 
of the species. 

\section{Extended BCS equations} 

At zero temperature the attractive interaction leads to pairing
of fermions which breaks the SU(3) symmetry but only two colors
are paired and one is left unpaired \cite{cinesi,baym,iome}. 
We assume, without loss of generality \cite{cinesi,baym,iome}, 
that the red and green particles are paired 
and the blue are not paired. The interacting terms can be then treated 
within the minimal mean-field BCS approximation, i.e. 
neglecting the Hartree terms while the pairing gap 
\beq 
\Delta=g \, \langle {\hat \psi}_{G}({\bf r},t){\hat \psi}_{R}({\bf r},t) 
\label{gappo}
\rangle
\eeq
between red and green fermions is the key quantity. 
In this way the mean-field Lagrangian density becomes 
\beq 
{\hat {\cal L}}_{mf} = 
\sum_{\alpha=R,G,B} {\hat \psi}^+_{\alpha} 
\left(i\hbar {\partial\over \partial t}
+{\hbar^2\over 2 m}\nabla^2 + \mu \right) {\hat \psi}_{\alpha} +
\Delta \, {\hat \psi}_{G}{\hat \psi}_{R} + 
\Delta \, {\hat \psi}_{G}^+{\hat \psi}_{R}^+  \; ,   
\label{ham-mf} 
\eeq
under the simplifying condition that the pairing gap is real, 
i.e. $\Delta^*=\Delta$. It is then straightforward to write down 
the Heisenberg equations of motion of the field operators:  
\beqa 
i\hbar {\partial \over \partial t} {\hat \psi}_R &=& 
\left( -{\hbar^2\over 2m}\nabla^2 - \mu \right) 
{\hat \psi}_R + \Delta \, {\hat \psi}_G^+ \; , 
\\
i\hbar {\partial \over \partial t} {\hat \psi}_G &=& 
\left( -{\hbar^2\over 2m}\nabla^2 - \mu \right) 
{\hat \psi}_G + \Delta \, {\hat \psi}_R^+ \; , 
\\
i\hbar {\partial \over \partial t} {\hat \psi}_B &=& 
\left( -{\hbar^2\over 2m}\nabla^2 - \mu \right) {\hat \psi}_B \; ,  
\eeqa
which are coupled by the presence of the same chemical potential $\mu$ 
in the three equations. 
We now use the Bogoliubov-Valatin representation of the 
field operator ${\hat \psi}_{\alpha}({\bf r},t)$ 
in terms of the anticommuting quasi-particle 
Bogoliubov operators ${\hat b}_{{\bf k}\alpha}$:  
\beqa 
{\hat \psi}_R({\bf r},t) &=& \sum_{\bf k} \left( 
u_k e^{i({\bf k}\cdot {\bf r}-\omega_k t)} \, {\hat b}_{{\bf k}R} - 
v_k e^{-i({\bf k}\cdot {\bf r}-\omega_k t)} \, {\hat b}_{{\bf k}G}^+ 
\right) \; , 
\\
{\hat \psi}_G({\bf r},t) &=& \sum_{\bf k} \left( 
u_k e^{i({\bf k}\cdot {\bf r}-\omega_k t)} \, {\hat b}_{{\bf k}G} +  
v_k e^{-i({\bf k}\cdot {\bf r}-\omega_k t)} \, 
{\hat b}_{{\bf k}R}^+ \right) \; , 
\\
{\hat \psi}_B({\bf r},t) &=& \sum_{\bf k} 
e^{i({\bf k}\cdot {\bf r}-\Omega_k t)} \, {\hat b}_{{\bf k}B} \; ,  
\eeqa
where $u_k$ and $v_k$ are such that $u_k^2+v_k^2=1$. 
After inserting these expressions into the Heisenberg equations of motion  
of the field operators we get 
\beqa 
\xi_k &=& \hbar \Omega_k = {\hbar^2k^2\over 2m}-\mu \; , 
\\
E_k &=& \hbar \omega_k = \sqrt{\xi_k^2 + \Delta^2} \; , 
\eeqa
and also 
\beq 
\label{vk}
u_k^2 = {1\over 2} \left( 1 + \frac{\xi_k}{E_k} \right) \; , 
\quad\quad\quad  
v_k^2 = {1\over 2} \left( 1 - \frac{\xi_k}{E_k} \right) \, .   
\eeq 
By imposing the following ground-state averages 
\beq 
\langle {\hat b}_{{\bf k}B}^+ {\hat b}_{{\bf k}'B} \rangle 
= \Theta(-\xi_k) \ \delta_{{\bf k},{\bf k}'} \; , 
\quad\quad 
\langle {\hat b}_{{\bf k}R}^+ {\hat b}_{{\bf k}R} \rangle 
= \langle {\hat b}_{{\bf k}G}^+ {\hat b}_{{\bf k}G} \rangle = 
\Theta(-E_k) \ \delta_{{\bf k},{\bf k}'} \; , 
\eeq
with $\Theta(x)$ the Heaviside step function, 
the number equation (\ref{numero}) gives 
\beq 
N = N_R + N_G + N_B \; , 
\label{bcs1} 
\eeq 
where 
\beq 
N_R = N_G = {1\over 2} \sum_{\bf k} v_k^2 
\label{bcs1a}
\eeq
and 
\beq 
N_B = \sum_{\bf k} \Theta\left(\mu-{\hbar^2k^2\over 2m}\right) 
\label{bcs1b}
\; .  
\eeq
Similarly, the gap equation (\ref{gappo}) gives 
\beq
-{1\over g} = {1 \over V} \sum_{\bf k} {1\over 2 E_k} \; . 
\label{bcsGap}
\eeq 
The chemical potential $\mu$ and the gap energy $\Delta$ are obtained by
solving equations (\ref{bcs1}) and (\ref{bcsGap}). 
In the continuum limit, due to the 
choice of a contact potential, the gap equation (\ref{bcsGap}) 
diverges in the ultraviolet. This divergence is 
linear in three dimensions and logarithmic in two dimensions. 
We shall face this problem in the next two sections. 

Another interesting quantity is the the number of red-green pairs 
in the lowest state, i.e. the condensate 
number of red-green pairs, that is given by 
\cite{sala-odlro,ortiz,ohashi1,ohashi2}
\beq
N_0 = \int d^3{\bf r}_1 \; d^3{\bf r}_2 \; | \langle 
{\hat \psi}_{G}({\bf r}_1)  
{\hat \psi}_{R}({\bf r}_2)  
\rangle |^2 = \sum_{\bf k} u_k^2 v_k^2 \; . 
\eeq 
In the last years two experimental groups \cite{zwierlein1,zwierlein2,ueda}
have analyzed the condensate fraction of three-dimansional ultra-cold 
two-hyperfine-component Fermi vapors of $^6$Li atoms in 
the crossover from the Bardeen-Cooper-Schrieffer (BCS) 
state of Cooper Fermi pairs to the Bose-Einstein condensate (BEC) 
of molecular dimers. These experiments are in quite good agreement 
with mean-field theoretical predictions 
\cite{sala-odlro,ortiz} and Monte-Carlo 
simulations \cite{astrakharchik} at zero temperature, 
while at finite temperature beyond-mean-field 
corrections are needed \cite{ohashi1,ohashi2}. 
Here we show how to calculate the condensate fraction $N_0/N$ for the 
three-component Fermi gas at zero temperature \cite{iome} 
in three \cite{sala-odlro} and two \cite{sala-odlro2} dimensions. 
Finally, we calculate the first sound velocity $c_s$ 
of the three-component 
system by using the zero-temperature thermodynamic formula \cite{landau2}
\beq 
c_s=\sqrt{{n\over m} \, {d\mu\over dn}} 
\label{cs-mio}
\eeq 
where $\mu$ is the chemical potential of the Fermi gas and 
$n$ the total density. The sound velocity $c_s$, which is the Nambu-Goldstone 
mode of pairing breaking of SU(3) symmetry, has been previously 
analyzed by He, Jin and Zhuang \cite{cinesi} in the three 
dimensional case. Here we study $c_s$ in the two dimensional case too. 

\section{Three dimensional case} 

In three dimensions a suitable regularization \cite{leggett,marini} 
of the gap equation (\ref{bcsGap}) is obtained by introducing 
the inter-atomic scattering length $a_F$ via the equation 
\beq
-{1\over g} = - {m \over 4 \pi \hbar^2 a_F} + 
{1 \over V} \sum_{\bf k} \frac{m}{\hbar^2k^2} \,,
\eeq 
and then subtracting this equation from 
the gap equation (\ref{bcsGap}). 
In this way one obtains the three-dimensional regularized gap equation 
\beq 
-{m \over 4 \pi \hbar^2 a_F} = {1 \over V} 
\sum_{\bf k} \left( { 1 \over 2 E_k} 
- \frac{m}{\hbar^2k^2} 
\right) . 
\label{bcs3d}  
\eeq 
In the three-dimensional continuum limit 
$\sum_{\bf k} \to V/(2\pi)^3 \int d^3{\bf k}
\to V/(2\pi^2) \int k^2 dk$ 
from  the number equation (\ref{bcs1}) with (\ref{bcs1a}) 
and (\ref{bcs1b}) we find the total number density as 
\beq  
n = {N\over V} = n_R + n_G + n_B \; , 
\eeq 
with 
\beq 
n_R = n_G = {1\over 2} {(2m)^{3/2} \over 2 \pi^2 \hbar^3} \,
\Delta^{3/2} \, I_2\!\left({\mu \over \Delta}\right)  \, ,
\label{gbcs2} 
\eeq 
and 
\beq 
n_B = {1\over 3} 
{(2m)^{3/2} \over 2 \pi^2 \hbar^3} \mu^{3/2} \ \Theta(\mu) \; . 
\eeq
The renormalized gap equation (\ref{bcs3d}) becomes instead 
\beq 
-{1\over a_F} = {2 (2m)^{1/2} \over \pi \hbar} \,
\Delta^{1/2} \, I_1\!\left({\mu \over \Delta}\right)  \, , 
\label{gbcs1} 
\eeq 
where $k_F=(6\pi N/(3V))^{1/3}=(2\pi^2n)^{1/3}$ is the Fermi wave number.  
Here $I_1(x)$ and $I_2(x)$ are the two monotonic 
functions 
\beq 
I_1(x) = \int_0^{+\infty} y^2 
\left( {1\over \sqrt{(y^2-x)^2+1}} - {1\over y^2} \right) dy \; , 
\eeq
\beq 
I_2(x) = \int_0^{+\infty} y^2 \left( 1 - {y^2-x\over \sqrt{(y^2-x)^2+1}}
\right) dy \; , 
\eeq
which can be expressed in terms of elliptic 
integrals, as shown by Marini, Pistolesi 
and Strinati \cite{marini}. In a similar way we get 
the condensate density of the red-green pair as 
\beq 
n_0 = {N_0\over V} = {m^{3/2} \over 8 \pi \hbar^3} \,
\Delta^{3/2} \sqrt{{\mu\over \Delta}+\sqrt{1+{\mu^2 \over \Delta^2} }} 
\label{con0} \; .   
\eeq 
This equation and the gap equation (\ref{gbcs1}) are the same 
of the two-component 
superfluid fermi gas (see \cite{sala-odlro}) but the number 
equation (\ref{bcs1}), with (\ref{bcs1a}) and (\ref{bcs1b}), is 
clearly different. Note that all the relevant quantities can be expressed 
in terms of the ratio 
\beq 
x_0 = {\mu\over \Delta} \; ,  
\eeq\
where $x_0\in ]-\infty,\infty[$. 
In this way the scaled energy gap $\Delta/\epsilon_F$ and the scaled 
chemical potential $\mu/\epsilon_F$ read 
\beq 
{\Delta\over\epsilon_F} = {x_0\over I_2(x_0)+{1\over 3} x_0^{3/2}
\Theta(x_0)} \; ,  
\eeq
\beq 
{\mu\over\epsilon_F} = {1\over I_2(x_0)+{1\over 3} x_0^{3/2} 
\Theta(x_0)} \; , 
\eeq
where $\epsilon_F=\hbar^2k_F^2/(2m)=(2\pi^2n)^{2/3}\hbar^2/(2m)$ 
is the Fermi energy of the 3D ideal three-component 
Fermi gas with total density $n$. 
The fraction of red fermions, which is equal to 
the fraction of green fermions, is given by 
\beq 
{n_R\over n} = {n_G\over n} = 
{I_2(x_0)\over 2 I_2(x_0) 
+ {2\over 3} x_0^{3/2}\ \Theta(x_0) } \; ,  
\eeq
while the fraction of blue fermions reads 
\beq 
{n_B\over n} = 1 - {I_2(x_0)\over I_2(x_0) 
+ {1\over 3} x_0^{3/2}\ \Theta(x_0) } \; . 
\eeq
The fraction of condensed red-green pairs is instead 
\beq 
{n_0\over n} = {\pi \over 8\sqrt{2}} 
{\sqrt{x_0+\sqrt{1+x_0^2}}\over I_2(x_0)+{1\over 3}
x_0^{3/2}\ \Theta(x_0) } \; . 
\eeq
Finally, the adimensional interaction strength of the BCS-BEC 
crossover is given by 
\beq 
y = {1\over k_F a_F} = -{2\over \pi} 
{I_1(x_0)\over \left(
I_2(x_0) + {1\over 3} x_0^{3/2}\ \Theta(x_0) \right)^{1/3} } \; . 
\eeq
We can use these parametric formulas of $x_0$ to plot the 
density fractions 
as a function of the scaled interaction strength $y$. 

\begin{figure}
\centerline{\psfig{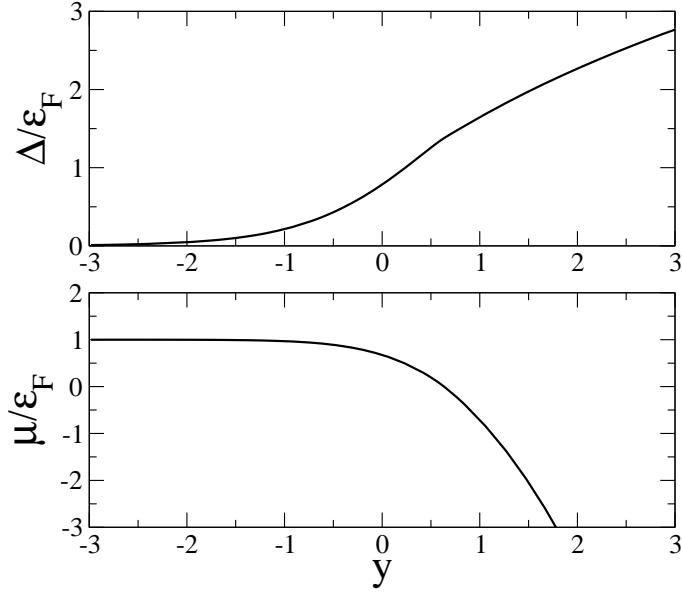}}
\small 
\caption{Ultracold fermions in three-dimensions. 
Upper panel: scaled energy gap $\Delta/\epsilon_F$ 
as a function of scaled interaction strength $y=1/(k_Fa_F)$. 
Lower panel: scaled chemical potential $\mu/\epsilon_F$ 
as a function of scaled interaction strength $y=1/(k_Fa_F)$.
Note that $k_F=(2\pi^2n)^{1/3}$ is the Fermi wave number and 
$\epsilon_F=\hbar^2k_F^2/(2m)=(2\pi^2n)^{2/3}\hbar^2/(2m)$ 
is the Fermi energy of the 3D ideal three-component 
Fermi gas with total 3D density $n$.} 
\label{fig1}
\end{figure} 

In the upper panel of Fig. \ref{fig1} we plot the energy gap 
$\Delta$ (in units of the Fermi energy $\epsilon_F$) 
as a function of scaled interaction strength $y=1/(k_Fa_F)$. 
As expected the gap $\Delta$ is exponentially small 
in the BCS region ($y\ll -1$), it becomes of the order 
of the Fermi energy $\epsilon_F$ at unitarity ($y=0$), 
and then it inceases in the BEC region ($y\gg 1$).  
In the lower panel of Fig. \ref{fig1} we show instead 
the scaled chemical potential $\mu/\epsilon_F$ 
as a function of scaled interaction strength $y=1/(k_Fa_F)$. 
In the BCS region ($y\ll -1$) the chemical potential $\mu$ 
is positive and practically equal to the Fermi energy $\epsilon_F$ 
of the ideal gas; at unitarity ($y=0$) the $\mu$ is 
still positive but close to zero; it becomes equal to zero 
at $y\simeq 0.6$ and then diminishes as $-y^2$ (half 
the binding energy of the formed dimers).

\begin{figure}
\centerline{\psfig{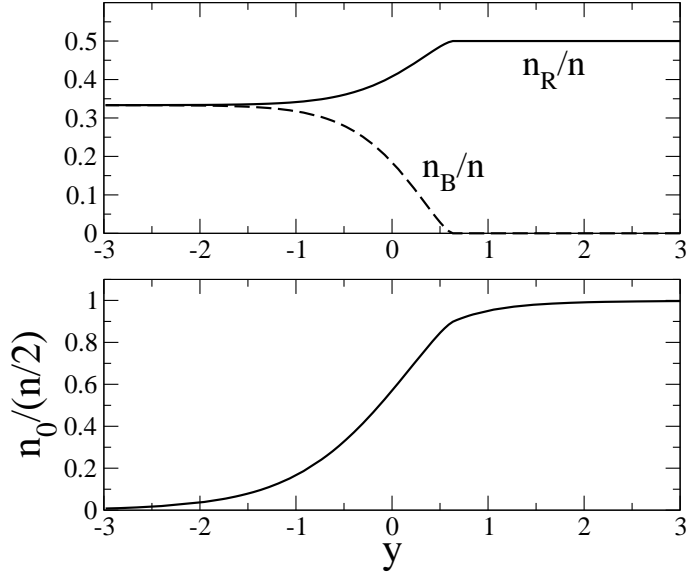}}
\small 
\caption{Ultracold fermions 
in three-dimensions. 
Upper panel: fraction of red fermions $n_R/n$ (solid line) 
and fraction of blue fermions $n_B/n$ (dashed line) 
as a function of scaled interaction strength $y=1/(k_Fa_F)$. 
Lower panel: condensed fraction of red-green particles $n_0/n$ 
as a function of scaled interaction strength $y=1/(k_Fa_F)$. Units as 
in Fig. 1.} 
\label{fig2}
\end{figure} 

In the upper panel of Fig. \ref{fig2} we plot the fraction 
of red fermions $n_R/n$ (solid line) and the 
fraction of blue fermions $n_B/n$ (dashed line) 
as a function of scaled interaction strength $y=1/(k_Fa_F)$. 
The behavior of $n_G/n$ is not shown because it is exactly 
the same of $n_R/n$. The figure shows that 
in the deep BCS regime ($y\ll -1$) 
the system has $n_R/n=n_G/n=n_B/n=1/3$. By increasing $y$ 
the fraction of red and green fermions increases while 
the fraction of blue fermions decreases. At $y\simeq 0.6$, 
where $\mu=0$, the fraction of blue fermions becomes zero, i.e. $n_B/n=0$ 
and consequently $n_R/n=n_G/n=1/2$. For larger 
values of $y$ there are only the paired red and green particles. 
This behavior is fully consistent with the findings of 
Ozawa and Baym \cite{baym}. In the lower panel of Fig.\ref{fig1} 
it is shown the plot of the condensate fraction $n_0/(n/2)$ of 
red-green pairs through the BCS-BEC crossover as a function of 
the Fermi-gas parameter $y=1/(k_Fa_F)$. 
The figure shows that a large condensate fraction builds up in the 
BCS side already before the unitarity limit ($y=0$), 
and that on the BEC side ($y\gg 1)$ it rapidly converges to one. 

\begin{figure}
\centerline{\psfig{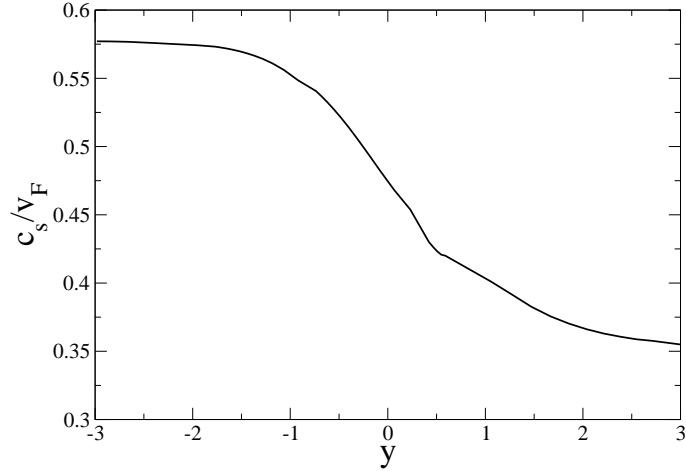}}
\small 
\caption{Ultracold fermions in three-dimensions. 
Scaled sound velocity $c_s/v_F$ 
as a function of scaled interaction strength $y=1/(k_Fa_F)$. 
Units as in Fig. 1, with $v_F=\hbar k_F/m$ the Fermi velocity of the 3D 
ideal three-component Fermi gas.} 
\label{fig3}
\end{figure} 

As previously stressed, by using Eq. (\ref{cs-mio}) one can obtain 
the first sound velocity. In particular, we have found that 
$\mu = \epsilon_F\ F(y)$, where $F(y)$ is the numerical function 
plotted in the lower panel of Fig. \ref{fig1}. It is then 
straightforward to show that 
\beq 
c_s = {v_F\over \sqrt{3}} \sqrt{F(y) - {1\over 2} y F'(y)} \; . 
\eeq
By using this formula we plot in Fig. \ref{fig3} the 
scaled sound velocity $c_s/v_F$ 
as a function of scaled interaction strength $y=1/(k_Fa_F)$. 
The curve shows that $c_s/v_F$ decreases by increasing $y$ and 
it shows a knee at $y\simeq 0.6$, where the chemical 
potential changes sign. 

\section{Two dimensional case} 

A two-dimensional Fermi gas can be obtained by imposing a very strong 
confinement along one of the three spatial directions. In practice, 
the potential energy $E_p$ of this strong external confinement 
must be much larger than the total chemical potential $\mu_{3D}$ 
of the fermionic system: $\mu_{3D}\ll 2 E_P$ \cite{io-me}. 
Contrary to the three-dimensional case, 
in two dimensions quite generally 
a bound-state energy $\epsilon_B$ exists for any value 
of the interaction strength $g$ between atoms \cite{randeria,marini}. 
For the contact potential the bound-state equation is 
\beq 
- {1 \over g} = 
{1 \over V} \sum_{\bf k} \frac{1}
{{\hbar^2k^2\over 2m} + \epsilon_B} \, ,
\eeq 
and then subtracting this equation from 
the gap equation (\ref{bcsGap}) 
one obtains the two-dimensional regularized gap equation 
\cite{randeria,marini} 
\beq 
\sum_{\bf k} \left( 
\frac{1}{ {\hbar^2k^2\over 2m} + \epsilon_B} 
- {1\over 2 E_k} \right) = 0 \; . 
\label{bcs2}  
\eeq 
Note that, for a 2D inter-atomic potential described 
by a 2D circularly symmetric well of radius $R_0$ and 
depth $U_0$, the bound-state energy $\epsilon_B$ is given 
by $\epsilon_B \simeq \hbar^2/(2mR_0^2) \exp{(-2\hbar^2/(mU_0R_0^2))}$ 
with $U_0 R_0^2 \to 0$ \cite{landau}.  

In the two-dimensional 
continuum limit $\sum_{\bf k} \to V/(2\pi)^2 \int d^2{\bf k}
\to V/(2\pi) \int k dk$, the Eq. (\ref{bcs2}) gives 
\beq 
\epsilon_B = \Delta 
\left( \sqrt{1+{\mu^2\over \Delta^2}}- {\mu\over \Delta} \right) \; .  
\label{gap-2d}
\eeq
Note that here $V$ is the 2D volume of the gas, i.e. an area. 
Instead, the number equation (\ref{bcs1}) with (\ref{bcs1a}) 
and (\ref{bcs1b}) gives the total number density as 
\beq 
n = {N\over V} = n_R + n_G + n_B \; ,
\eeq 
where $V$ is a two-dimensional volume (i.e. an area), the red and green  
densities are 
\beq 
n_R = n_G = {1\over 2} \big({m \over 2\pi\hbar^2}\big) \Delta 
\left( {\mu\over \Delta} + \sqrt{1+{\mu^2\over \Delta^2}} \right) \; ,  
\label{numb} 
\eeq
while the blue density reads 
\beq 
n_B = \big({m \over 2\pi\hbar^2}\big) \mu \ \Theta(\mu) \; . 
\eeq
Finally, the condensate density of red-green pairs is given by 
\beq 
n_0 = {1\over 4} \big({m\over 2 \pi \hbar^2}\big) \Delta 
\left( {\pi\over 2} + \arctan{({\mu\over \Delta})} \right) \; . 
\label{cond}
\eeq 

Also in this two-dimensional case all the relevant quantities can be expressed 
in terms of the ratio $x_0 = {\mu/\Delta}$, 
where $x_0\in ]-\infty,\infty[$. In particular, 
the scaled pairing gap is given by 
\beq 
{\Delta\over \epsilon_F} = {3\over x_0+\sqrt{1+x_0^2}+x_0\Theta(x_0)} \; , 
\eeq
while the scaled chemical potential reads 
\beq 
{\mu\over \epsilon_F} = {3x_0\over x_0+\sqrt{1+x_0^2}+x_0\Theta(x_0)} \; ,  
\eeq 
where the two-dimensional Fermi energy $\epsilon_F=\hbar^2k_F^2/(2m)$ 
of the 2D ideal three-component Fermi gas with 2D total density $n$ 
is given by $\epsilon_F=\pi \hbar^2 n/m$ with $k_F=(4\pi n/3)^{1/2}$ 
the Fermi wave number. 

The fraction of red fermions, which is equal to 
the fraction of green fermions, is given by 
\beq 
{n_R\over n} = {n_G\over n} = 
{x_0+\sqrt{1+x_0^2}\over 2 [x_0 + \sqrt{1+x_0^2}+x_0 \ \Theta(x_0) ] }  \; , 
\eeq 
the fraction of blue fermions is 
\beq 
{n_B\over n} = 1 - 2 {n_R\over n} \; , 
\eeq
and the condensate fraction is 
\beq 
{n_0\over n} = { {\pi\over 2} + \arctan{(x_0)} \over 
4[ x_0 + \sqrt{1+x_0^2} + x_0 \ \Theta(x_0) ] }  \; .   
\eeq 
It is convenient to express the bound-state energy $\epsilon_B$ 
in terms of the Fermi energy $\epsilon_F$. In this way we find 
\beq 
{\epsilon_B\over \epsilon_F} = 
3 { 
\sqrt{1+ x_0^2} - x_0   
\over 
x_0+ \sqrt{1+ x_0^2} + x_0\ \Theta(x_0) } \; ,  
\label{epsilonB} 
\eeq
We can now use these parametric formulas of $x_0$ to plot the fractions 
as a function of the scaled bound-state energy $\epsilon_B/\epsilon_F$. 

\begin{figure}
\centerline{\psfig{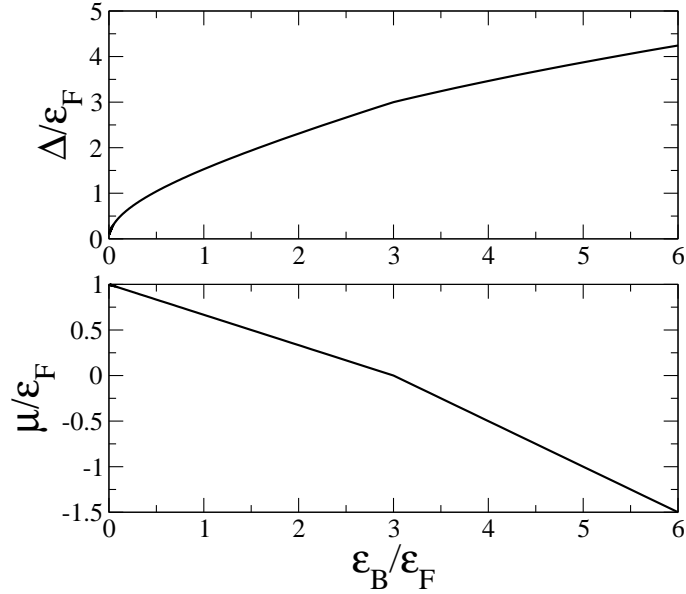}}
\small 
\caption{Ultracold 
fermions in two-dimensions. 
Upper panel: scaled energy gap $\Delta/\epsilon_F$ 
as a function of scaled bound-state energy $\epsilon_B/\epsilon_F$. 
Lower panel: scaled chemical potential $\mu/\epsilon_F$ 
as a function of scaled bound-state energy $\epsilon_B/\epsilon_F$.
Note that $k_F=(4\pi n/3)^{1/2}$ is the Fermi wave number and 
$\epsilon_F=\hbar^2k_F^2/(2m)=(4\pi n/3) \hbar^2/(2m)$ 
is the Fermi energy of the 2D ideal three-component 
Fermi gas with total 2D density $n$.} 
\label{fig4}
\end{figure} 

In the upper panel of Fig. \ref{fig4} we plot the scaled energy gap 
$\Delta/\epsilon_F$ 
as a function of scaled binding energy $\epsilon_B/\epsilon_F$. 
The gap $\Delta$ is extremely small 
in the ``BCS region'' ($\epsilon_B/\epsilon_F\ll 1/2$), 
it becomes of the order of the Fermi energy $\epsilon_F$ at 
$\epsilon_B/\epsilon_F=1/2$, and then it inceases 
in the ``BEC region'' ($\epsilon_B/\epsilon_F\gg 1/2$).  
In the lower panel of Fig. \ref{fig4} we show instead 
the scaled chemical potential $\mu/\epsilon_F$ 
as a function of scaled binding energy $\epsilon_B/\epsilon_F$. 
In the BCS region ($\epsilon_B/\epsilon_F\ll 1/2$) 
the chemical potential $\mu$ 
is positive and decreases as $\mu = \epsilon_F - \epsilon_B/3$;  
$\mu$ becomes equal to zero 
at $\epsilon_B/\epsilon_F = 3$ and then it further decreases linearly 
as $\mu = (3 \epsilon_F - \epsilon_B)/2$. 

\begin{figure}
\centerline{\psfig{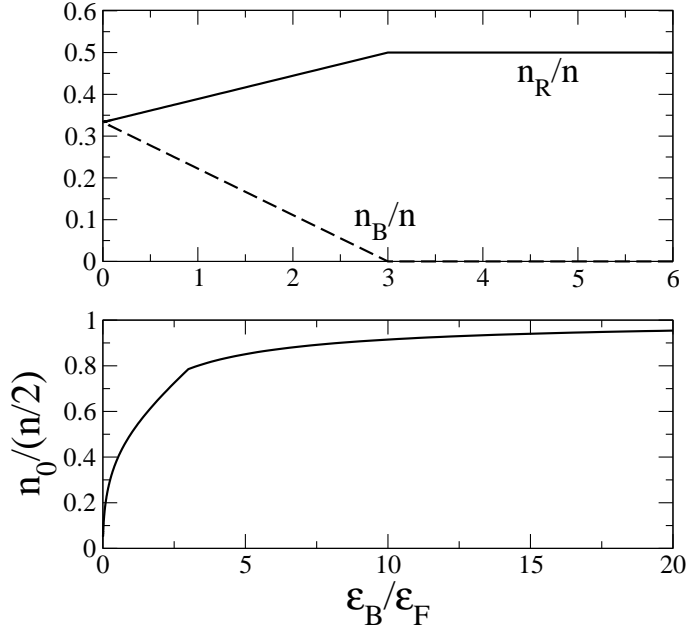}}
\small 
\caption{Ultracold fermions in two-dimensions. 
Upper panel: fraction of red fermions $n_R/n$ (solid line) 
and fraction of blue fermions $n_B/n$ (dashed line) 
as a function of scaled bound-state energy $\epsilon_B/\epsilon_F$. 
Lower panel: condensed fraction of red-green particles $n_0/n$ 
as a function of scaled bound-state energy $\epsilon_B/\epsilon_F$. 
Units as in Fig. 4.} 
\label{fig5}
\end{figure} 

In the upper panel of Fig. \ref{fig5} we plot the fraction 
of red fermions $n_R/n$ (solid line) and the 
fraction of blue fermions $n_B/n$ (dashed line) 
as a function of scaled bound-state energy $\epsilon_B/\epsilon_F$.
The behavior of $n_G/n$ is not shown because it is exactly 
the same of $n_R/n$. The figure shows that 
in the deep BCS regime ($\epsilon_B/\epsilon_F \ll 1$) 
the system has $n_R/n=n_G/n=N_B/n=1/3$. 
By increasing $\epsilon_B/\epsilon_F$ 
the fraction of red and green fermions increases while 
the fraction of blue fermions decreases. At $\epsilon_B/\epsilon_F
= 3$, where $\mu=0$, the fraction of blue fermions becomes zero. For larger 
values of $\epsilon_B/\epsilon_F$ there are only 
the paired red and green particles. 
This behavior is quite similar to the one of the 
three-dimensional case; the main difference is due to the fact 
that here the curves are linear. 
In the lower panel of Fig.\ref{fig5} 
it is shown the condensate fraction $n_0/(n/2)$ of 
red-green pairs. 
In the weakly-bound BCS regime ($\epsilon_B/\epsilon_F \ll 1$) 
the condensed fraction $n_0/n$ goes to zero, while in the 
strongly-bound BEC regime ($\epsilon_B/\epsilon_F \gg 1$) 
the condensed fraction $n_0/n$ goes to $1/2$, i.e. all the 
red-green Fermi pairs belong to the Bose-Einstein condensate. 
Notice that the condensate fraction is zero 
when the bound-state energy $\epsilon_B$ is zero. 
For small values of $\epsilon_B/\epsilon_F$ the condensed 
fraction has a very fast grow but then it reaches the 
asymptotic value $1/2$ very slowly. 

\begin{figure}
\centerline{\psfig{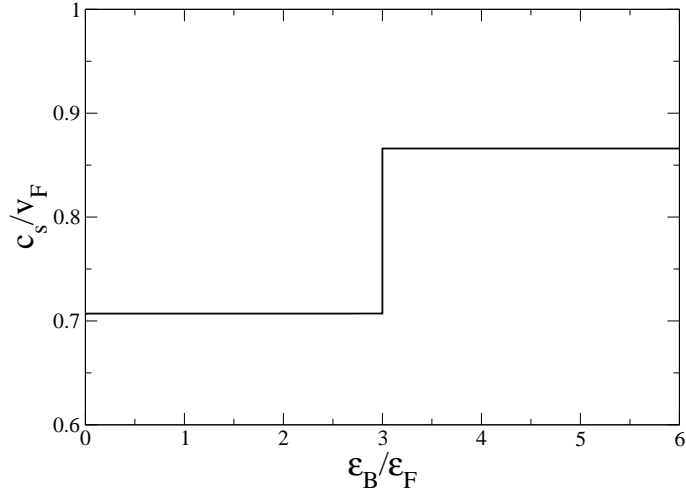}}
\small 
\caption{Ultracold fermions in two-dimensions. 
Scaled sound velocity $c_s/v_F$ 
as a function of scaled bound-state energy $\epsilon_B/\epsilon_F$. 
Units as in Fig. 4, with $v_F=\hbar k_F/m$ the Fermi velocity of the 2D 
ideal three-component Fermi gas.} 
\label{fig6}
\end{figure} 

Also in 2D, by using Eq. (\ref{cs-mio}) one can obtain 
the first sound velocity. We have found that 
$\mu = \epsilon_F\ G(\epsilon_B/\epsilon_F)$, 
where $G(\epsilon_B/\epsilon_F)$ is the numerical function 
plotted in the lower panel of Fig. \ref{fig4}. It is then 
straightforward to show that 
\beq 
c_s = {v_F\over \sqrt{2}} \sqrt{ G({\epsilon_B\over \epsilon_F}) 
- {\epsilon_B\over\epsilon_F} 
G'({\epsilon_B\over \epsilon_F})} \; . 
\eeq
By using this formula we plot in Fig. \ref{fig6} the 
scaled sound velocity $c_s/v_F$ 
as a function of scaled binding energy $\epsilon_B/\epsilon_F$. 
The curve shows that $c_s/v_F$ is constant, i.e. 
$c_s/v_F = 1/\sqrt{2}$, by increasing 
$\epsilon_B/\epsilon_F$ up to $\epsilon_B/\epsilon_F=3$ 
where the chemical potential becomes equal to zero. 
For a larger value of $\epsilon_B/\epsilon_F$ the sound velocity 
$c_s$ jumps to a larger constant value, i.e. $c_s/v_F = \sqrt{3}/2$. 
This kink in the first sound velocity $c_s$ is reminescent 
of the jumps seen with repulsive fermions with reduced  
dimensionalities \cite{io-me,flavio} and with dipolar interaction \cite{das}.

\section{Conclusions}

We have investigated a uniform three-component ultracold 
fermions by increasing 
the SU(3) invariant attractive interaction. We have considered 
the symmetry breaking of the SU(3) symmetry due to the formation 
of Cooper pairs both in the three-dimensional case and in the 
two-dimensional one. We have obtained explicit formulas 
and plots for energy gap, chemical potential, 
number densities, condensate density, population imbalance 
and first sound velocity in the full BCS-BEC crossover. 
In our calculations we have used the zero-temperature 
mean-field extended BCS theory, which is expected to give reliable results 
apart in the deep BEC regime \cite{astrakharchik,ohashi1,ohashi2}. 
Our results are of interest for next future experiments 
with degenerate gases made of alkali-metal or alkaline-earth atoms. 
As stressed in the introduction, SU(N) invariant interactions 
can be experimentaly obtained by using these atomic 
species \cite{selim,hara,ana}. The problem of unequal couplings, 
and also that of a fixed number of atoms for each component, 
is clearly of big interest too, and its analysis can be 
afforded by including more than one order parameter \cite{torma}. 

There are other interesting open problems about superfluid ultracold atoms 
we want to face in the next future. In particular, 
we plan to investigate quasi one-dimensional and quasi 
two-dimensional Bose-Einstein condensates 
in nonlinear lattices (i.e. with space-dependent interaction strength)  
\cite{boris-nl}. 
Moreover, we want to analyze the signatures of classical and quantum 
chaos \cite{sala-chaos1,sala-chaos2,sala-chaos2bis,sala-chaos3,sala-chaos4} 
with Bose-Einstein condensates in single-well and 
double-well configurations, and also in the presence of vortices 
\cite{sala-chaos5,sala-chaos6,sala-chaos7}. Finally, 
we aim to calculate analytically the coupling tunneling energy 
of bosons by means of the WKB semiclassical quantization 
\cite{sala-wkb1,sala-wkb2,sala-wkb3,sala-wkb4} and comparing 
it with the numerical results of the Gross-Pitaevskii 
equation. 

The author thanks Luca Dell'Anna, Giovanni Mazzarella, 
Nicola Manini, Carlos Sa de Melo, Flavio Toigo 
and Andrea Trombettoni for useful discussions and suggestions.


\begin{thebibliography}{99}

\bibitem{book1} A.J. Leggett, Quantum Liquids: 
Bose Condensation and Cooper Pairing in 
Condensed-Matter Systems (Oxford Univ. Press, Oxford, 2006). 

\bibitem{book2} H.T.C. Stoof, B.M. Dennis, and K. Gubbels, 
Ultracold Quantum Fields (Springer, Berlin, 2009). 

\bibitem{sigh07a} G. Diana, N. Manini, and L. Salasnich, 
Phys. Rev. A {\bf 73} 065601 (2006). 

\bibitem{sigh07b} L. Salasnich and N. Manini, 
Laser Phys. {\bf 17}, 169 (2007). 

\bibitem{sigh10a} L. Salasnich, F. Ancilotto, and F. Toigo, 
Laser Phys. Lett. {\bf 7}, 78 (2010).

\bibitem{sigh10b} Y. Cheng and S.K. Adhikari, 
Laser Phys. Lett. {\bf 7}, 824 (2010). 

\bibitem{sigh10c} Y.I. Yukalov, Laser Phys. Lett. {\bf 7}, 467 (2010). 

\bibitem{sigh11} R.F. Shiozaki, G.D. Telles, 
Y.I. Yukalov, and V.S. Bagnato, Laser Phys. Lett. {\bf 8}, 393 (2011).

\bibitem{selim} T.B. Ottenstein, T. Lompe, M. Kohen, 
A.N. Wenz, and S. Jochim, Phys. Rev. Lett. {\bf 101}, 
203202 (2008). 

\bibitem{hara} J.H. Huckans, J.R. Williams, E.L. Hazlett, R.W. Stites, 
and K. M. O'Hara, Phys. Rev. Lett. {\bf 102}, 165302 (2009). 

\bibitem{wu1} C. Wu, J.P. Hu, and S.C. Zhang, 
Phys. Rev. Lett. {\bf 91}, 186402 (2003).

\bibitem{wu2} C. Wu, Mod. Phys. Lett. B {\bf 20}, 1707 (2006). 

\bibitem{ana} A.V. Gorshkov, M. Hermele, V. Gurarie, 
C. Xu, P.S. Julienne, J. Ye, P. Zoller, E. Demler, 
M.D. Lukin, A.M. Rey, Nature Physics {\bf 6}, 289 (2010). 

\bibitem{various1} A.G.K. Modawi and A.J. Leggett, 
J. Low Temp. Phys. {\bf 109}, 625 (1997). 

\bibitem{various2} C. Honerkamp and W. Hofstetter, Phys. Rev. Lett. 
{\bf 92}, 17040 (2004). 

\bibitem{various3} T. Paananen, J.-P. Martikainen, 
and P. Torma, Phys. Rev. A {\bf 73}, 053606 (2006). 

\bibitem{various4} C.K. Chung and C.K. Law, 
Phys. Rev. A {\bf 82}, 033620 (2010).

\bibitem{cinesi} L. He, M Jin, and P. Zhang, 
Phys. Rev. A {\bf 74}, 033604 (2006).

\bibitem{baym} T. Ozawa and G. Baym, Phys. Rev. A {\bf 82}, 
063615 (2010). 

\bibitem{eagles} D.M. Eagles, Phys. Rev. {\bf 186}, 456 (1969). 

\bibitem{leggett} A.J. Leggett, in {\it Modern Trends in the Theory 
of Condensed Matter}, 
p. 13, edited by A. Pekalski and J. Przystawa (Springer, Berlin, 1980). 

\bibitem{randeria} M. Randeria, J.-M. Duan, and L.-Y. Sheih, 
Phys. Rev. B {\bf 41}, 327 (1990). 

\bibitem{marini} M. Marini, F. Pistolesi, and G.C. Strinati,
Eur. Phys. J. B {\bf 1}, 151 (1998).

\bibitem{sigh08} M.Y. Kagan and S.L. Ogarkov, 
Laser Phys. {\bf 18}, 509 (2008).

\bibitem{iome} L. Salasnich, Phys. Rev. A {\bf 83}, 033630 (2011). 

\bibitem{zwierlein1} M.W. Zwierlein, C.A. Stan, C.H. Schunck, 
S.M.F. Raupach, A.J. Kerman, and W. Ketterle, 
Phys. Rev. Lett. {\bf 92}, 120403 (2004). 

\bibitem{zwierlein2} M.W. Zwierlein, C.H. Schunck, C.A. Stan, 
S.M.F. Raupach, and W. Ketterle, 
Phys. Rev. Lett. {\bf 94}, 180401 (2005).  

\bibitem{ueda} Y. Inada, M. Horikoshi, S. Nakajima, 
M. Kuwata-Gonokami, M. Ueda, and T. Makaiyama, 
Phys. Rev. Lett. {\bf 101}, 180406 (2008). 

\bibitem{sala-odlro} L. Salasnich, N. Manini, and A. Parola, 
Phys. Rev. A {\bf 72}, 023621 (2005).  

\bibitem{ortiz} G. Ortiz and J. Dukelsky, 
Phys. Rev. A {\bf 72}, 043611 (2005). 

\bibitem{astrakharchik} G. E. Astrakharchik,
J. Boronat, J. Casulleras, and S. Giorgini,
Phys. Rev. Lett. {\bf 95}, 230405 (2005).

\bibitem{ohashi1} Y. Ohashi and A. Griffin, 
Phys. Rev. A {\bf 72}, 063606 (2005). 

\bibitem{ohashi2} N. Fukushima, Y. Ohashi, E. Taylor, and A. Griffin, 
Phys. Rev. A {\bf 75}, 033609 (2007). 

\bibitem{sala-odlro2} L. Salasnich, Phys. Rev. A {\bf 76}, 
015601 (2007). 

\bibitem{landau2} L.D. Landau and E.M. Lifshits, {\it 
Statistical Physics}, Part 2, vol. 9 
(Butterworth-Heinemann, Oxford, 1980).  

\bibitem{landau} L.D. Landau and E.M. Lifshitz, 
{\it Quantum Mechanics. Non Relativistic Theory. 
Course of Theoretical Physics}, 
Vol. 3 (Pergamon Press, New York, 1989).

\bibitem{io-me} G. Mazzarella, L. Salasnich, and F. Toigo, 
Phys. Rev. A {\bf 79}, 023615 (2009). 

\bibitem{flavio} L. Salasnich, and F. Toigo, 
J. Low Temp. Phys. {\bf 150}, 643 (2008).

\bibitem{das} J.P. Kestner and S. Das Sarma, 
Phys. Rev. A {\bf 82}, 033608 (2010). 

\bibitem{torma} O.H.T. Nummi, J.J. Kinnunen, 
and P. Torma, New J. Phys. {\bf 13}, 055013 (2011).

\bibitem{boris-nl} Y.V. Kartashov, B.A. Malomed, 
and L. Torner, Rev. Mod. Phys. {\bf 83}, 247 (2011). 

\bibitem{sala-chaos1} L. Salasnich, Phys. Rev. D {\bf 52}, 6189 (1995). 

\bibitem{sala-chaos2} L. Salasnich, Mod. Phys. Lett. A {\bf 12}, 1473 (1997). 

\bibitem{sala-chaos2bis} L. Salasnich, Phys. Lett. A {\bf 266}, 187 (2000).

\bibitem{sala-chaos3} A.R. Kolovsky and A. Buchleitner, 
Europhys. Lett. {\bf 68} 632 (2004). 

\bibitem{sala-chaos4} C. Weiss and N. Teichmann, 
Phys. Rev. Lett. 100, 140408 (2008).  

\bibitem{sala-chaos5} L. Salasnich, Int. J. Mod. Phys. B {\bf 14}, 1 (2000). 

\bibitem{sala-chaos6} L. Salasnich, Laser Phys. {\bf 14}, 291 (2004). 

\bibitem{sala-chaos7} S.K. Adhikari and L. Salasnich, 
Phys. Rev. A {\bf 75}, 053603 (2007). 

\bibitem{sala-wkb1} M. Robnik and L. Salasnich, 
J. Phys. A: Math. Gen. {\bf 30}, 1711 (1997). 

\bibitem{sala-wkb2} M. Robnik and L. Salasnich, 
J. Phys. A: Math. Gen. {\bf 30}, 1719 (1997). 

\bibitem{sala-wkb3} G. Alvarez, J. Math. Phys. {\bf 45}, 3095 (2004). 

\bibitem{sala-wkb4} A.V. Turbiner, Int. J. Mod. Phys. A {\bf 25}, 647 (2010).

\end{thebibliography}
\end{document}